\documentclass[11pt]{article}

\usepackage[english]{babel}
\usepackage{graphicx}
\usepackage{color,amssymb,color}
\usepackage[bookmarks=true,bookmarksopen=true,colorlinks=true,breaklinks=true,linkcolor=blue,citecolor=blue]{hyperref}
\usepackage{amsmath,amsfonts,amssymb,mathrsfs,amsthm}
\usepackage[affil-it,max2]{authblk}
\usepackage{enumerate}   
\usepackage{multicol}

\usepackage[margin=3.0cm]{geometry}

\newcommand{\Eqref}[1]{Eq.~\eqref{#1}}
\newcommand{\Eqsref}[1]{Eqs.~\eqref{#1}}
\newcommand{\Sectionref}[1]{section~\ref{#1}}

\title{Approximations of the quasi-local Bartnik mass in general relativity}

\author[1]{Leon Escobar-Diaz}
\author[2]{Chris Stevens}
\affil[1]{Department of Mathematics, Universidad del Valle, Colombia.}
\affil[2]{School of Mathematics and Statistics, University of Canterbury, New Zealand.}

\begin{document}
\maketitle

\begin{abstract}
In this study, we employ eth-operators and spin-weighted spherical harmonics to express the ADM mass of a static space-time based on the mean values of its components over a radius-$r$ sphere. While initially derived for standard spherical coordinates, we showcase its adaptability by demonstrating its usefulness in expressing a quasilocal mass —specifically, the Bartnik mass— of an almost round 2D-hypersurface in terms of some specific boundary conditions. Additionally, we utilize this formulation to propose a deep learning methodology for numerically constructing static metrics that incorporate 2D-hypersurfaces with specified Bartnik masses.
\end{abstract}

\section{Introduction}\label{introduction}

The necessity for a well-defined and computable definition of local mass within the realm of general relativity is directly motivated by the fundamental physics imperative to establish a connection with classical Newtonian gravity. However, unlike many other scientific disciplines, the concept of mass in general relativity is notably unintuitive and intricate. The primary reason for this lies in the fact that the energy carried by the gravitational field is not included in the energy–momentum tensor. Consequently, establishing a clear and appropriate concept of energy within a localized region of spacetime, and consequently defining a notion of local mass, becomes a challenging task. This issue is widely recognized in the scientific community as the \textit{quasi-local mass problem}.

Numerous proposals aiming to define quasi-local mass in general relativity have emerged in recent decades. Notable among these are Hawking's mass, Penrose's mass, Geroch's mass, and Bartnik's mass (refer to \cite{szabados2009quasi} for an overview). Bartnik's mass, in particular, has garnered significant attention from the scientific community in recent years. This heightened interest is attributed to its foundation on the well-established concept of the ADM-mass, recognized as a positive and invariant quantity of the spacetime \cite{bartnik1986mass,schoen1979proof, witten1981new}.

In general terms, Bartnik's mass is defined as the mass of a local spatial region within the spacetime, delineated by a surface boundary $\Sigma$. It is quantified as the supremum of the set comprising all ADM-masses derived from every possible asymptotically flat three-dimensional manifold that isometrically encompasses the region enclosed by $\Sigma$. To be precise, let $\mathcal{AF}$ represent the space containing all $(M, g_{ })$ three-dimensional asymptotically flat manifolds meeting the assumptions of the positive mass theorem (refer to \cite{schoen1979proof}) and containing a bounded region with surface boundary $\Sigma$. Then, the mass contained inside the region internally bounded by $\Sigma$ is given by 
\begin{equation}\label{minimizer}
 M_{B}(\Sigma):= \text{inf} \{ M_{ \text{ADM} }( g_{ }) | \ (M, g_{ }) \in \mathcal{AF}  \}.
\end{equation}
Throughout this work, Greek characters such as $\mu$, $\nu$, etc., will denote tensor components relative to a specific frame. Coordinate frame vectors will be represented by $\mathbf{e}_{\mu}$ and their corresponding covector by $\mathbf{w}^{\mu}$. Additionally, we will utilize Einstein's summation convention (for further details, refer to \cite{Wald:1984un}).

In \cite{corvino2000scalar}, it was demonstrated that the infimum can only be achieved by a static metric, provided suitable boundary conditions are imposed on $\Sigma$. These solutions are commonly referred to as \textit{static metric extensions} of $\Sigma$ in the literature. Consequently, the problem of discovering a smooth minimizer of \Eqref{minimizer} is essentially reduced to identifying appropriate static metric extensions of $\Sigma$.

The subsequent inquiry pertains to determining the necessary boundary conditions on $\Sigma$ that ensure the existence of solutions to the static vacuum Einstein equations. This query prompted Bartnik to introduce his well-known \textit{static metric extension conjecture} in \cite{bartnik1997energy}. This conjecture, also recognized in the literature as Bartnik's conjecture, requires a clear comprehension of the mathematical properties defining a static metric before formally stating it. Following \cite{Wald:1984un}, a spacetime is called static if it possesses a timelike killing vector, indicating that its geometry remains unchanged under time translations. Denoting by $t$ the temporal coordinate,  this property allows a static spacetime metric $g^{(4)}_{ }$ to be expressed using the line element:
\begin{eqnarray}\label{static_metric}
  g^{(4)}_{ } = - f^2 dt  \otimes dt  +  g_{ }  ,
\end{eqnarray}
where $dt$ are the covectors asociated to the temporal killing vector $\partial_t$ with norm $f$, and $g_{ }$ represents the induced metric on the three-dimensional surfaces $M$ orthogonal to it. Furthermore, a static must satisfy the vacuum Einstein field equations 
\begin{eqnarray*}
\mathbf{Ric} - \dfrac{ \mathbf{R} }{2} \ g^{(4)}_{ } = 0,
\end{eqnarray*}
where $\mathbf{Ric}$ is the Ricci tensor  and $\mathbf{R}$ the scalar curvature. Thus,
exploiting the temporal symmetry of the spacetime, one can perform a symmetry reduction (refer, for instance, to the appendix of \cite{geroch1971method}) of the equations along the temporal killing vector to derive the static Einstein equation:
\begin{equation}\label{SEE}
\begin{aligned}
\Delta_{ g} f & = 0,\\
f R_{\mu \nu } & = \nabla_\mu \nabla_\nu f,
\end{aligned}
\end{equation}
where $R_{\mu \nu }$ are the components of the Ricci tensor of the 3-dimensional induced metric $g$, and $\nabla$  and  $\Delta_{g}$ are the covariant derivative and Laplacian compatible with it, respectively. When expressed in coordinates, this system of tensorial equations transforms into a coupled system of partial differential equations (PDEs). The solutions of these equations constitute the components of the metric $g_{ }$ and the norm $f$, forming a pair $(g_{ },f)$ that determines the static metric \Eqref{static_metric}.  For instance, in standard spherical coordinates $(r,\theta,\phi)$ (with respect to the coordinate frame), the flat 3D metric $\mathring{g}_{ }$ and the constant function $f=1$ constitute a pair $(\mathring{g}_{ },\dot{f})$ satisfying the aforementioned system. This pair corresponds to the Minkowski metric:
\begin{equation*}
g^{(4)}_{ } = - dt  \otimes dt  + \mathring g .
\end{equation*}

In the general case, solving the system of equations \Eqsref{SEE} requires prescribing certain information or constraints about the desired solution at the boundary of the domain. Bartnik's conjecture posits that a surface $\Sigma$ with metric $\gamma_{ }$ and a positive function $H$ can serve as a means to impose boundary conditions on the system of partial differential equations. This allows for the determination of a pair $(g_{ },f)$ representing a static extension of the region $\Sigma$ with metric $\gamma_{ }$ and mean curvature $H$. Mathematically, this reduces to a boundary value problem, seeking the pair $(g_{ },f)$ such that $\Sigma \subset  M$ and:
\begin{equation}\label{EE}
\left.  \begin{aligned}
   \Delta_{ g} f    &=0 \\
   f R_{ \mu \nu}         &= \nabla_{\mu} \nabla_{\nu} f  
\end{aligned}
\right\} \text{ in }  M,
\end{equation}
and 
\begin{equation}\label{BC}
\left.  \begin{aligned}
    g    &=  \gamma  \\
   \mathcal{H}     &=  H 
\end{aligned}
\right\} \text{ on }  \Sigma,
\end{equation}
where $\mathcal{H}:=-\nabla_{\nu} n^{\nu}$, with $n$ being the normal vector to the surface $\Sigma$. In the literature, these last two boundary conditions are commonly known  as \textit{Bartnik's data}. 

In his pioneering work \cite{miao2003existence}, Miao established Bartnik's conjecture for a specific class of Bartnik data. He achieved this by assuming $\gamma_{ }$ and $H$ as perturbations, possessing $\mathbb{Z}_2\times\mathbb{Z}_2\times\mathbb{Z}_2$ symmetry, of the unit sphere and its mean curvature in flat space, respectively. Miao proved the existence of static extensions in which both the components of the metric $g$ and the norm $f$ are sufficiently close, within a weighted Sobolev norm (for a formal definition, refer to \cite{bartnik1986mass}), to the flat metric. Years later, Anderson in \cite{anderson2015local}, extended Miao's results by removing the symmetry assumptions. Consequently, the general result of the existence and uniqueness of static perturbations of the flat metric with Bartnik data sufficiently close to the unit sphere in the flat manifold was established. 

Following Anderson's work, Wiygul in \cite{wiygul2018bartnik}, employed a similar approach to Miao, implementing harmonic coordinates. Wiygul provided an estimate of the ADM mass, allowing for the approximation of the Bartnik mass of surfaces $\Sigma$ with Bartnik data as linear perturbations of the unit sphere in the flat manifold by means of a simple formula. Moreover, recent efforts (\cite{harvie2022mass,wiygul2021second}) have been made to generalize that formula to accommodate non-linear perturbations. For a comprehensive and up-to-date review on the subject, refer to \cite{anderson2023bartnik}.

In this work, we utilize eth-operators and spin-weighted spherical harmonics (as detailed in \cite{beyer2014numerical}) to express the $M_{ADM}(g)$ of a static space-time $g$ in terms of the mean values of its components over a sphere of radius $r$. While this expression is originally derived for standard spherical coordinates, we demonstrate its broader applicability in expressing the Bartnik mass $M_{B}(\Sigma)$ of linear perturbations on the unit sphere within a flat manifold. This expression is formulated in terms of the mean values of the Bartnik data $\gamma_{ }$ and $H$, both defined on the boundary $\Sigma$. Furthermore, we establish its consistency with Wiygul's estimate \cite{wiygul2018bartnik}.

Additionally, we leverage this formulation of $M_{ADM}(g)$ to propose a neural network approach for numerically constructing static metrics that incorporate $\Sigma$ with specified Bartnik masses. Through this numerical exploration, we aim to shed light on potential numerical methodologies for addressing this problem. Another intriguing numerical proposal addressing a similar problem, based on an inverse mean curvature geometric flow and assuming axial symmetry, is discussed in \cite{Cederbaum2019AFA}.

This paper is organized as follows: in \Sectionref{sec:1}, we present a brief overview of the mathematical background concerning the eth-operators and spin-weight spherical harmonics. Later, in \Sectionref{sec:2}, we utilize the eth-formalism to express the $M_{ADM}(g)$ of a general metric $g$ in terms of the mean value of the metric components within a sphere of radius $r$. This expression will be further utilized in \Sectionref{sec:3} to derive a simple equation, consistent with Wiygul's estimate, for computing the Bartnik mass $M_{B}(\Sigma)$ of linear perturbations of the unit sphere in the flat manifold. This computation will be based on the mean values of the Bartnik data $\gamma_{ }$ and $H$ defined on the boundary $\Sigma$. Moreover, in \Sectionref{sec:4}, we introduce a neural network methodology aimed at numerically constructing static metrics with specified Bartnik masses. Finally, in \Sectionref{sec:5}, we provide a discussion on the main contributions of this work.

\section{Mathematical preliminaries for the eth-operators}\label{sec:1}

\subsection{The spin-weighted spherical harmonics and the eth-operators}\label{sec:swsh}

To begin with, let us consider a square integrable complex function on the 2-sphere of radius one  $\mathbb{S}^2(1)$, i.e, $f: \mathbb{S}^2(1) \to \mathbb{C}$. For simplicity we will impose standard spherical coordinates $(\theta,\varphi)$ on $\mathbb{S}^2(1)$. Following Penrose and Rindler \cite{penrose1984spinors}, we say that $f$ has spin-weight $s$ if it transforms under the action of the one-parameter group $U(1)$ in the tangent plane at every point $(\theta,\varphi) \in \mathbb{S}^2(1)$  as $f \to e^{\text{i} s \zeta} f$, where $\zeta$ is the group parameter. Furthermore, $f$ can be written as 
\begin{equation}\label{ec:spectral_decomposition0}
f = \sum\limits_{l,m}  \hspace{0.1cm}_{s}a_{lm} \hspace{0.2cm}_{s}Y_{lm} (\theta,\varphi) :=  \sum\limits_{l=|s|}^{\infty}  \sum\limits_{m=-l}^{l} \hspace{0.1cm}_{s}a_{lm} \hspace{0.2cm}_{s}Y_{lm} (\theta,\varphi) , 
\end{equation}
where the $_{s}Y_{lm}( \theta , \varphi)$ functions are the \textit{spin-weighted spherical harmonics} (SWSH), which are given in terms of the Legendre polynomials. The  coefficients ${}_{s}a_{lm}$ are called  \textit{the spectral coefficients}, and in analogy to the Fourier series, we will refer to ${}_{0}a_{00}$ as \textit{the fundamental mode}. Note that if the spin-weight of $f$ is $s>0$, the fundamental mode ${}_{0}a_{00}$ and all the coefficients of other spin-weights vanish. For further details regarding these basis functions and their mathematical properties, additional insights can be found in references such as \cite{alcubierre2008introduction}, which elaborates on the properties and applications of these spin-weighted spherical harmonics.

An important relation of the SWSH is their relation of orthogonality, given by the following relation 
\begin{eqnarray}\label{integral_properties_spherical_harmonics}
  \langle \  {}_{s_1} Y_{l_1 m_1 }(\theta,\varphi) \ , \ _{s_2}\overline{Y}_{l_2 m_2}(\theta,\varphi) \ \rangle   &=& \int \limits_{\mathbb{S}^2(1)}  {}_{s_1} Y_{l_1 m_1 }(\theta,\varphi) \: {}_{s_2}\overline{Y}_{l_2 m_2}(\theta,\varphi)   \sin \theta d\theta d\varphi ,\nonumber \\
   &=& \delta_{l_1 l_2} \delta_{m_1 m_2} \delta_{s_1 s_2}.
\end{eqnarray}

Note that $\langle \ , \ \rangle $ is just the standard inner product of $L^2 (\mathbb{S}^2(1)(1))$. Clearly, the above relation implies that the SWSH are an orthogonal basis for any function $f$ with spin-weight $s$ defined on $\mathbb{S}^2(1)$. From this, and using the fact that ${}_{0}Y_{00} = 1/\sqrt{4\pi}$, we can compute the mean value of a function $f \in \mathbb{S}^2(1)$ by
\begin{eqnarray}\label{mean_value_functions}
   \overline{f} &:=&
   \dfrac{1}{4 \pi}  \int \limits_{\mathbb{S}^2(1)} \  f \:   \ \sin \theta d\theta d\varphi = \dfrac{1}{ \sqrt{4\pi} }  \int \limits_{\mathbb{S}^2(1)} \  f \:  {}_{0} \overline{Y}_{0 0} \sin \theta d\theta d\varphi \nonumber, \\ 
   & = & \dfrac{1}{\sqrt{4\pi}}  \int \limits_{\mathbb{S}^2(1)} \  \left(   \sum\limits_{l,m}  \hspace{0.1cm}_{s}a_{lm} \hspace{0.2cm}_{s}Y_{lm} (\theta,\varphi)  \right)  {}_{0}\overline{Y}_{0 0}  \ \sin \theta d\theta d\varphi, \\
   &=& \dfrac{ {}_{0} a_{00} }{\sqrt{4\pi}} 
   = \dfrac{ \langle f , \ {}_{0}Y_{00} \rangle }{\sqrt{4 \pi}} ,\nonumber
\end{eqnarray}
which clearly implies that the mean value of $f$ on $\mathbb{S}^2(1)$ is proportional to its fundamental mode. \\

The \textit{eth-operators}, denoted by $\eth$ and $\bar{\eth}$, are  defined as (see for instance \cite{newman1966note}) 
\begin{equation}\label{eq:def_eths}
\eth f       := \partial_\theta f - \dfrac{\text{i}}{ \text{sin} \theta} \partial_\varphi f- s f \text{cot} \theta, \quad 
\bar{\eth} f := \partial_\theta f + \dfrac{\text{i}}{ \text{sin} \theta} \partial_\varphi f + s f \text{cot} \theta .
\end{equation}
These two operators \textit{raise} and \textit{lower} the spin-weights of the SWSH basis by means of the following two properties 
\begin{eqnarray}
\begin{aligned}\label{eq:eths}
\eth  \hspace{0.1cm}_{s}Y_{lm} (\theta,\varphi)  &= S(l,s, \hspace{0.3cm} 1) \hspace{0.1cm}_{s+1}Y_{lm} (\theta,\varphi) ,  \\
\bar{\eth}   \hspace{0.1cm}_{s}Y_{lm} (\theta,\varphi)   &= S(l,s, -1) \hspace{0.1cm}_{s-1}Y_{lm} (\theta,\varphi) , \\
\end{aligned}
\end{eqnarray}
where 
\begin{equation*}\label{ec:ecuationforS}
S(l,s, \zeta) := -\zeta \sqrt{ (l -\zeta s)(l+\zeta s+1) } \ .
\end{equation*} 
Additionally, combining the raising and lower properties, it follows that
\begin{eqnarray*}\label{eq:laplacian_eths}
    \bar{\eth} \eth  \hspace{0.1cm}_{s}Y_{lm} (\theta,\varphi) =   \eth \bar{\eth}  \hspace{0.1cm}_{s}Y_{lm} (\theta,\varphi) = -l(l+1) \hspace{0.1cm}_{s}Y_{lm} (\theta,\varphi). 
\end{eqnarray*}
An important consequence of the above statement is that if a function $f$ has a spin-weight of $s$ with spectral coefficients ${}_{s} a_{lm}$, then the application of the $\eth$ operator to $f$, denoted as $\eth f$, will result in a spin-weight of $s+1$ with spectral coefficients given by ${}_{s+1} a_{lm} = {}_{s} a_{lm}$. Similarly, applying the $\bar{\eth}$ operator to $f$, denoted as $\bar{\eth} f$, will yield a spin-weight of $s-1$ with spectral coefficients ${}_{s-1} a_{lm} = {}_{s} a_{lm}$. From this, it follows that if a function $f$ has a spin-weight of $1$, its fundamental mode should be zero. Consequently, upon lowering its spin-weight by applying $\bar{\eth}$, the fundamental mode remains zero. To illustrate this, consider the function $f$ with a spin-weight of $1$, written as:
\begin{equation*}
f =   \sum\limits_{l=1}^{\infty}  \sum\limits_{m=-l}^{l} \hspace{0.1cm}_{1}a_{lm} \hspace{0.2cm}_{1}Y_{lm} (\theta,\varphi).
\end{equation*}
Then, applying the operator  $\bar\eth$ we obtain
\begin{equation*}
\begin{aligned}
\bar \eth f &=   \sum\limits_{l=1}^{\infty}  \sum\limits_{m=-l}^{l} \hspace{0.1cm}_{1}a_{lm}  \ \bar \eth   \hspace{0.2cm}_{1}Y_{lm} (\theta,\varphi),   \\
 &= \sum\limits_{l=1}^{\infty}  \sum\limits_{m=-l}^{l} \hspace{0.1cm}_{1}a_{lm}   \ S(l,1,   -1 )  \hspace{0.2cm}_{0}Y_{lm} (\theta,\varphi),\\ 
\end{aligned}
\end{equation*}
which we can write as
\begin{equation}\label{ec:spectral_decomposition}
 \bar \eth f  = \sum\limits_{l=0}^{\infty}  \sum\limits_{m=-l}^{l} \hspace{0.1cm}_{0}\tilde{a}_{lm}   \ S(l,1,   -1 )  \hspace{0.2cm}_{0}Y_{lm} (\theta,\varphi),
\end{equation}
with ${}_{0} \tilde{a}_{lm}:={}_{1} a_{lm} S(l,1,   -1 )$ and ${}_{0} \tilde{a}_{00}:=0$. A similar result can be obtained if $f$ has a spin-weight of $-1$ as we raise its spin by applying $\eth$.

\subsubsection{The non-coordinate smooth frame adapted to the SWSH }\label{sec:coordinates}

Let us choose the \textit{non-coordinate frame} $(\mathbf{e}_1, \mathbf{e}_2)$ on some $S \simeq \mathbb{S}^2(1) $ as\footnote{This frame is extensively used in the well known Newman--Penrose formalism for gravitational waves, see for example \cite{alcubierre2008introduction}.}
\begin{equation}\label{eq:referenceframe}
\quad \mathbf{e}_{1}:=\frac 1{\sqrt 2}\left(\partial_{\theta}-\frac{\text{i}} {\sin\theta}\partial_\varphi\right), \quad \mathbf{e}_{2}  :=\frac 1{\sqrt 2}\left(\partial_{\theta}+\frac{\text{i}} {\sin\theta}\partial_\varphi\right),
\end{equation}
where $\partial_{\theta}$ and $\partial_{\varphi}$ correspond to the coordinate vectors associated to the standard spherical coordinates. Additionally, we define the coframe $(\mathbf{w}^1, \mathbf{w}^2)$  such that 
\begin{equation}\label{eq:referencecoframe}
\mathbf{w}^1 := \frac 1{\sqrt 2}\left( d \theta_a + \text{i} \sin \theta \ d \varphi_a \right), \quad 
\mathbf{w}^2 := \frac 1{\sqrt 2}\left( d \theta_a - \text{i} \sin \theta \ d \varphi_a \right),
\end{equation} 

where $d$ represents the exterior derivative operator in $\mathbb{S}^2(1)$ (refer to \cite{nakahara2018geometry}), where $\mathbf{w}^\nu \mathbf{e}_{\mu} = \delta^{\nu}_{\mu}$. It can be readily demonstrated that the frame vectors undergo a transformation when subjected to a rotation by an angle $\zeta$ in the tangent plane at each point of $\mathbb{S}^2(1)$. This transformation adheres to the action of the $U(1)$ group (as discussed in, for instance, \cite{beyer2016numerical}) as

\begin{equation}\label{eq:transformationrules_spin0}
 \mathbf{e}_\mu  \to  e^{\text{i} \tilde \Omega_\mu \zeta} \mathbf{e}_\mu , \text{ with } \tilde \Omega_\mu = \left\{
	\begin{array}{ll}
		\hspace{0.3cm}  1  & \mbox{if } \mu=1 ,\\
		-1 & \mbox{if } \mu=2 .
	\end{array}
\right.
\end{equation}

Since scalar numbers must be invariant under rotation, it follows from \Eqsref{eq:referencecoframe} that  the coframe   must transform under the action of U(1) as $\mathbf{w}_\mu  \to  e^{- \text{i} \tilde \Omega_\mu \zeta} \mathbf{w}_\mu$.  Consequently, tensor components with respect that frame (and coframe) must transform  as  
\begin{equation*}
T^{\mu...\nu}_{\qquad  \sigma...\lambda} \to e^{\text{i} s \zeta} T^{\mu...\nu}_{\qquad  \sigma...\lambda},
\end{equation*}
where the spin-weight $s$ depends on the number of frame vectors $\mathbf{e}_{\mu}$ and coframe covectors $\mathbf{w}^\mu$ with respect to the tensor components that are taken. Consequently, tensor components possess a clearly defined spin-weight.  For example, the components $\gamma_{\mu\nu}$ of the metric tensor $\gamma_{ }$ transform as
\begin{eqnarray*}\label{spines_metrica}
\gamma_{\mu\nu} =  \gamma (\mathbf{e}_\mu , \mathbf{e}_\nu) \to  \gamma(
 e^{\text{i} \tilde \Omega_\mu \zeta} \mathbf{e}_\mu ,  e^{\text{i} \tilde \Omega_\nu \zeta} \mathbf{e}_\nu)    ) = e^{\text{i} (\tilde \Omega_\mu+\tilde \Omega_\nu) \zeta} \gamma (\mathbf{e}_\mu , \mathbf{e}_\nu) = e^{\text{i} (\tilde \Omega_\mu+\tilde \Omega_\nu) \zeta} \gamma_{\mu \nu} . 
\end{eqnarray*}
From this, we deduce that their spin-weight $s$ is determined by $\tilde \Omega_\mu + \tilde \Omega_\nu$. Furthermore, from \Eqref{ec:spectral_decomposition0} it follows that  tensor components with spin-weight $s$ on $S \simeq \mathbb{S}^2(1)$ can be written as  
\begin{equation}\label{eq:TensorfunctionS2}
T^{\mu... }{}_{\sigma... } =  \sum\limits_{l,m}  \hspace{0.1cm}_{s}a_{lm} \hspace{0.2cm}_{s}Y_{lm} (\theta,\varphi).
\end{equation}


In order to compute the covariant derivatives on tensors and obtain the Ricci tensor, it is imperative to take into account the commutator for the non-coordinate smooth frame. This commutator, denoted as $[\mathbf{e}_\mu, \mathbf{e}_\nu]$, is defined by the structure constants $C^{\sigma}_{\mu \nu}$  as
\begin{equation*}
[\mathbf{e}_\mu, \mathbf{e}_\mu]  = C^{\sigma}_{\mu \nu} \mathbf{e}_\sigma.
\end{equation*}
It is worth noting that we classify the frame ${\mathbf{e}_{\mu}}$ as coordinate when all the $C^{\sigma}_{\mu \nu}$ vanish. In general, the components of the connection coefficients are given in the non-coordinate frame as (see for instance \cite{stephani2009exact})
\begin{equation*}\label{connection_coeffieicnrts}
 \Gamma^{\alpha}_{\beta \gamma}=\dfrac{1}{2}g^{\alpha \sigma}(  \partial_\gamma g_{\sigma \beta}+ \partial_\beta g_{\sigma \gamma}- \partial_\sigma g_{\beta \gamma} +  C_{\sigma \beta\gamma}+C_{\sigma \gamma\beta}-C_{\beta \gamma \sigma}   ),
\end{equation*}
where we have used $C^{\alpha} {}_{\mu \nu}:= g^{\alpha \rho} C_{\rho \mu \nu}$. From here,  the covariant derivatives of tensors can be expressed by
\begin{equation}\label{cov.derive}
 \nabla_{\tau} T^{\mu...}{}_{\nu...} =  \mathbf{e}_\tau \left(  T^{\mu ...}{}_{\nu ...} \right) - \Gamma^{\mu}_{\tau \sigma }  T^{\sigma...}{}_{\nu...} ... + \Gamma^{\sigma}_{\tau \nu}  T^{\mu...}{}_{\sigma...} ... \quad ,
\end{equation}
where second and third terms on the right-hand side denote the contraction of connection coefficients with the contravariant and covariant indices of the tensor, respectively. Assuming that the tensor component $s$ has some spin-weight $s$, we can compute the action of the frame vector over it in terms of the eth-operators by combining \Eqref{eq:def_eths} with the coordinate definition of the frame vectors \Eqsref{eq:referenceframe}, which yields
\begin{eqnarray}\label{eq:ethm}
 \mathbf{e}_\tau ( T^{\mu... }{}_{\sigma... }  ) = \dfrac{1}{\sqrt 2} \left(  \eth_\tau T^{\mu... }{}_{\sigma... }  +  \Omega_\tau  \ s \ \cot\theta  \  T^{\mu... }{}_{\sigma... }  \right),
\end{eqnarray}
where $\eth_{1} = \eth$ and $\eth_{2} = \bar \eth$. 
Finally, the components of Riemann tensor are given by 
\begin{equation*}\label{Riemanntensor}
R^{\mu} {}_{\nu \alpha \beta}=\Gamma^{\mu}_{\nu \beta, \alpha}-\Gamma^{\mu}_{\nu \alpha, \beta} +\Gamma^{\rho}_{\nu \beta}\Gamma^{\mu}_{\rho \alpha}-\Gamma^{\rho}_{ \nu \alpha}\Gamma^{\mu}_{\hspace{0.2cm} \rho \beta}-C^{\rho}_{\alpha \beta}\Gamma^{\mu}_{ \nu \rho} \;.
\end{equation*}

Consequently, from the aforementioned considerations, we deduce that by projecting tensors onto $S$ within the non-coordinate smooth frame $\mathbf{e}_\mu$, one can express their components in terms of the SWSH. Hence, the computation of covariant derivatives can be achieved by utilizing the eth-operators. This approach is commonly referred to as the \textit{eth-formalism} (see, for instance, \cite{beyer2016numerical} and references therein).

\section{The ADM mass and the spin-weighted spherical harmonics}\label{sec:2}

\subsection{The extended non-coordinate frame}
In this section, we derive a formula for $M_{ADM}(g)$ in terms of the mean values of the metric components by employing spin-weighted spherical harmonics. To do so, we consider the three dimensional manifold $(M,g_{ })$ covered by the standard spherical coordinates $(r,\theta,\varphi)$. 
Let us extend the non-coordinate smooth frame of the previous section to $(\mathbf{e}_0 ,\mathbf{e}_1, \mathbf{e}_2)$ such that $\mathbf{e}_0 := \partial_r$ and $\mathbf{e}_1$, $\mathbf{e}_2$ are the frame vectors  defined  in \Eqref{eq:referenceframe}. Then, we can compute the connection coefficients, covariant derivatives and Ricci tensor components by following the same procedure 
discussed in the previous section by noting that since $\mathbf{e}_0$ is orthogonal to $\mathbb{S}^2(1)$, it is invariant under the group of rotation U(1). Then the extended non-coordinate smooth frame transform is
\begin{equation}\label{eq:transformationrules_spin}
 \mathbf{e}_\mu  \to  e^{\text{i} \Omega_\mu \zeta} \mathbf{e}_\mu , \text{ with } \Omega_\mu = \left\{
	\begin{array}{ll}
        \hspace{0.3cm}  0  & \mbox{if } \mu=0 ,\\   
		\hspace{0.3cm}  1  & \mbox{if } \mu=1 ,\\
		-1 & \mbox{if } \mu=2 .
	\end{array}
\right.
\end{equation}
Thus, by the discussion of the previous section, we can write the components of tensor in terms of the SWSH and thus we can use the \textit{eth-formalism} in the extended frame $\mathbf{e}_\mu$. Note that for consistency we use
\begin{eqnarray*} 
 \mathbf{e}_0 ( T^{\mu... }{}_{\sigma... }  ) = \dfrac{1}{\sqrt 2} \left(  \partial_r T^{\mu... }{}_{\sigma... }     \right),
\end{eqnarray*}
while $\mathbf{e}_1 ( T^{\mu... }{}_{\sigma... }  )$ and $\mathbf{e}_2 ( T^{\mu... }{}_{\sigma... }  )$ are computed as in \Eqref{eq:ethm}. For a general 3-dimensional metric $g_{ }$ with components in a non-coordinate frame denoted by $g_{\mu\nu}$, we extend the discussion from \Sectionref{sec:coordinates} and utilize the transformation detailed in \Eqref{eq:transformationrules_spin} to express the metric components in terms of SWSH as
\begin{equation}\label{eq:spectral_metric}
 g_{\mu\nu} := \sum^{\infty}_{l=0} \sum^{l}_{m=-l}   {}_{ (\Omega_{\mu}+\Omega_{\nu}) }  \hat{g}_{\mu\nu}(r)_{lm}  \quad { }_{ (\Omega_{\mu}+\Omega_{\nu})} Y_{lm}(\theta,\phi),
\end{equation}
where ${}_{  (\Omega_{\mu}+\Omega_{\nu}) }\hat{g}_{\mu\nu}(r)_{lm}$ are the spectral coefficients of the metric components $g_{\mu\nu}$ with spin-weight given by $ (\Omega_{\mu}+\Omega_{\nu})$. As a particular case of this, let us assume  that $\Sigma$ is a flat manifold. Then,  
in the smooth frame the flat metric takes the following form 
\begin{eqnarray*}
     \mathring{g}_{ } =  \mathbf{w}^0  \otimes \mathbf{w}^0  + 2 r^2 \mathbf{w}^1 \otimes  \mathbf{w}^2  ,
\end{eqnarray*}
where we have the non-coordinate coframe $(\mathbf{w}^0 ,\mathbf{w}^1,\mathbf{w}^2)$, with  $\mathbf{w}^1$ and  $\mathbf{w}^2$ being the coframe covectors defined in \Eqref{eq:referencecoframe}, and $\mathbf{w}^0  := d r  $. The components of the metric can be written in matricial form as  
\begin{equation}\label{eq:metric_form}
  \mathring g_{ } =  
 \begin{pmatrix}
1 & 0 & 0\\
0 & 0 & r^2 \\
0 & r^2 & 0 \\
\end{pmatrix}.
\end{equation}
From here we can clearly see that the non-vanishing metric components have spin-weight zero and their components can be written in terms of ${ }_{0} Y_{00}(\theta,\phi)$.

\subsection{The ADM mass in the extended non-coordinate frame}
Next we consider the ADM mass, $M_{ADM}(g)$, of the metric $g_{ }$ in this frame. 
In order to simplify the computations and to obtain a simple formula for $M_{ADM}(g)$, we will take the components corresponding to $\mu =1$ and $\nu=2$ scaled with a factor of $r^2$, i.e., we take $ g(\mathbf{e}_1 , \mathbf{e}_2 ):= g_{12} r^{2}$.  From \cite{gourgoulhon20123+}, we write expression for  $M_{ADM}(g)$  as 

\begin{eqnarray*}
   M_{ADM}(g) = \dfrac{1}{16 \pi} \lim_{S \to \infty} \oint_{S}  \left(
   \mathring{ g}^{\sigma \lambda} \mathring{ \nabla}_{\sigma}  g_{\rho \lambda} -\mathring{ \nabla}_{\rho}( \mathring{ g}^{\nu \mu}  g_{\nu \mu}  ) \right) \  n^{\rho} d A    ,
\end{eqnarray*}
where $\mathring\nabla_{\rho}$ is the covariant derivative compatible with the flat metric $\mathring{g}_{ }$, $ n^{\rho}$ are the components of unit normal outward-directed vector to the closed 2-dimensional hypersurface $S$, and $dA$ is its corresponding area element in $S$. Since in principle  $S$ is arbitrary, we can choose $S = \mathbb{S}^2(r)$, which leads to $n^{\rho} := (1,0,0)$ and $dA = r^2 \sin \theta d\theta d\phi$ in standard spherical coordinates. Thus, by defining the vector integrand as
\begin{eqnarray}\label{eq:integrandd}
  \mathcal{I}_\rho:= \mathring{ g}^{\sigma \lambda} \mathring{ \nabla}_{\sigma}  g_{\rho \lambda} -\mathring{ \nabla}_{\rho}( \mathring{ g}^{\nu \mu}  g_{\nu \mu}  ),
\end{eqnarray}
we can write  
\begin{eqnarray}\label{M_ADM_expression1}
   M_{ADM}(g) = \dfrac{1}{8 \sqrt{\pi} } \ \lim_{r \to \infty} \   \oint_{\mathbb{S}^2(1)}                     \mathcal{I}_0  \  {}_{0}Y_{00}(\theta,\phi)   \ r^2 d \Omega^2 ,
\end{eqnarray}
where $d\Omega^2 := d\theta^2 + \sin^2\theta d\phi^2$ and we have used the fact ${}_{0}\overline{Y}_{00}(\theta,\phi)  = 1/ \sqrt{4 \pi} $. As discussed in \Sectionref{sec:coordinates}, since we are using the non-coordinate smooth frame, the components of the vector integrand must have a well defined spin-weight. Thus, we can write the component $\mathcal{I}_0$ as
\begin{equation*}
    \mathcal{I}_0 = \sum^{\infty}_{l=0} \sum^{l}_{m=-l}  {}_0 \mathbb{I}(r)_{lm}  \ {}_0 Y_{lm}(\theta,\phi),
\end{equation*}
where ${}_0 \mathbb{I}(r)_{lm}$ are the spectral coefficients with vanishing spin-weight. Substituting this expression into \Eqref{M_ADM_expression1}, and using the orthogonality of the SWSH  (\Eqref{integral_properties_spherical_harmonics}), we obtain 
\begin{eqnarray}\label{eq:Mass_equation}
   M_{ADM}(g) = \dfrac{1}{4 \sqrt{4 \pi} } \ \lim_{r \to \infty} \ {}_0 \mathbb{I},(r)_{00} \ r^2 =   \dfrac{1}{4 } \ \lim_{r \to \infty} \ \overline{ \mathcal{I} }_{0} \ r^2.
\end{eqnarray}
 Using the relation between the covariant derivative and the eth-operators with the flat metric in the non-coordinate smooth frame, we obtain after some computation that \Eqref{eq:integrandd} can be written
\begin{eqnarray*}
    \mathcal{I}_{0} &=& \mathring{ g}^{\sigma \lambda} \mathring{ \nabla}_{\sigma}  g_{0 \lambda} -\mathring{ \nabla}_{0}( \mathring{ g}^{\nu \mu}  g_{\nu \mu}  ),  \\ &=&  \frac{\eth g_{02}}{\sqrt{2} r^2} +\frac{\bar{\eth }   g_{01}}{\sqrt{2} r^2} + \frac{2}{r} \left(   g_{00} - g_{12} \right) - 2 \partial_r g_{12} \ .
\end{eqnarray*}
Note that because $g_{01}$ and $g_{02}$ have spin $1$ and $-1$ respectively, the fundamental mode of $\bar{\eth } g_{01} $ and $\eth g_{02}$  is zero (see \Eqref{ec:spectral_decomposition}). Hence, the mean value (see \Eqref{mean_value_functions}) of $\mathcal{I}_{0}$ on $\mathbb{S}^{2}(1)$ is given by
\begin{equation*}
       \overline{ \mathcal{I} }_{0} = \dfrac{1}{\sqrt{4 \pi}} \langle \mathcal{I}_{0} \ , \ {}_{0} Y_{00} (\theta,\phi) \rangle  = \frac{2}{r} \left(   \overline g_{00} - \overline g_{12} \right) - 2 \partial_r \overline g_{12}.
\end{equation*}
Finally, substituting the above into \Eqref{eq:Mass_equation} leads to the expression
\begin{eqnarray}\label{eq:Mass_equation_final}
   M_{ADM}(g) =  \frac{1}{2}  \ \lim_{r \to \infty} \  r\Big{[} \left(   \overline g_{00} - \overline g_{12} \right) -   r \partial_r \overline g_{12}\Big{]} \ .
\end{eqnarray}
As a result, we have obtained an expression for $M_{ADM}(g)$ of  in terms of its mean values on the 2-sphere. Additionally, note that this limits exist and is possibly non-vanishing if and only if $\overline{g}_{00}$ and $\overline{g}_{12}$  are  $\mathcal{O}(r^{-1})$.

\section{The Bartnik Mass of an almost round 2D-hypersurface}\label{sec:3}

\subsection{The equations for the mean values}
In this section we will find the Bartnik mas of an ``almost round" 2D-hypersurface surface  $\Sigma$ with metric and mean curvature given respectively by
\begin{eqnarray}\label{bartnik_data}
   \gamma_{ } := \mathring \gamma_{ } + \epsilon  \gamma^{(\epsilon)}_{ }, \qquad H = \mathring H + \epsilon  H^{(\epsilon)}, 
\end{eqnarray}
where $\epsilon$ is some small constant and $\mathring \gamma_{ }$ and $\mathring H$ are the standard metric and mean curvature of the 2-sphere of radius $r_0$, which we denote by $\mathbb{S}^{2}(r_0)$ in the flat metric. In the smooth frame $\mathring \gamma_{ }$ is written as $\mathring \gamma_{ }=  r^2 \   \mathbf{w}^1 \otimes  \mathbf{w}^2$, while  the scalar function $\mathring H$ is just $2/r$. 
By ``almost round" hypersurfaces we mean that we assume that   $\epsilon^2  \approx 0$. i.e., we assume a surface $\Sigma$ that can be considered as linear perturbation of the metric and mean curvature of $\mathbb{S}^{2}(r_0)$. \\

As it was discussed in \Sectionref{introduction}, in order to compute the Bartnik  mass of $\Sigma$, we  have to solve the the static system  (\ref{EE}-\ref{BC}) for the pair $(g_{ },f)$. Therefore, we will search for solutions of the form 
\begin{eqnarray}\label{form_of_perturbations}
g_{ } := \mathring{g}_{ } + \epsilon g^{(\epsilon)}_{ }, \quad f := 1 + \epsilon u ,
\end{eqnarray}
i.e., $g_{ }$ will be a linear perturbation of the flat metric. Because of \Eqref{eq:Mass_equation_final}, we know that only the metric components with spin-weight zero contribute to $M_{ADM}$. Thus, we assume that the perturbation $g^{(\epsilon)}_{ } $ has the form
\begin{equation}\label{perturbationlinearregime}
      g^{(\epsilon)}_{ } :=   h \left( \ \mathbf{w}^0 \otimes  \mathbf{w}^0  +  2   r^2    \mathbf{w}^1 \otimes  \mathbf{w}^2  \right).
\end{equation}
Note that the coframe  $(\mathbf{w}^1,\mathbf{w}^2)$ defines the metric components of a 2-dimensional surface which has unit normal vector $n^{\mu}=( ( 1+ \epsilon h )^{-1/2} , 0, 0) \approx (  1 - \epsilon  h /2 , 0, 0)$. Then, it follows that the mean curvature of this 2-dimensional surface is 
\begin{eqnarray}\label{mean_curvature_1}
\mathcal{H} = -\nabla_{\mu} n^{\mu} = -\dfrac{2}{r} +  \frac{h}{r} -  \partial_r  h .
\end{eqnarray}

Since we found that the ADM mass only depends on the mean values of the metric components (and its derivatives) (\Eqref{eq:Mass_equation_final}), and considering that only the components of the perturbation  $g^{(\epsilon)}_{ }$ can contribute to $M_{ADM}$, we will search for the fundamental modes of $h$ and $u$ that satisfy the  static system \Eqref{EE}, \Eqref{BC}. Therefore, we will solve the system

\begin{equation}\label{LEE}
\left.  \begin{aligned}
\overline{ \delta( f R_{\mu \nu }) }  &=&    \overline{ \delta( \nabla_{\mu} \nabla_{\nu } f ) }   \\
 \overline{  \delta (  \Delta_{ g} f  ) }  & =& 0  \hspace{1.4cm}
\end{aligned}
\right\} \text{ in }  M,
\end{equation}
with the boundary conditions
\begin{equation}\label{LBC}
\left.  \begin{aligned}
 \overline{g}^{(\epsilon)}_{ }    &=    \overline{\gamma}^{(\epsilon)}_{ }   \\
  \overline{\mathcal{H}}^{(\epsilon)}      &=   \overline{H}^{(\epsilon)}  
\end{aligned}
\right\} \text{ on }  \mathbb{S}^2(r_0),
\end{equation}
where we have used the notation
\begin{eqnarray*}
    \delta q := \dfrac{d q}{d \epsilon} \Big|_{\epsilon=0},
\end{eqnarray*}
to denote the variation with respect to the $\epsilon$ parameter, and the overline in the equations to denote the inner product $ \overline{T}_{\mu \nu}=\langle T_{\mu \nu} , \ {}_{0}Y_{00}(\theta,\phi) \rangle /\sqrt{4 \pi}$ for tensor components $T_{\mu \nu}$,  according to \Eqref{mean_value_functions}. From now on we will refert to the system \Eqsref{LEE} as the \textit{linearized static equations}.

\subsection{Solution of the static system for the mean values of the metric}
One can obtain from the equation for the fundamental mode of $f$
 \begin{eqnarray*}
       \overline{ \delta( \Delta_{g} f ) }   &=& 0, \\
     \partial_{rr} \overline{u} + \dfrac{2}{r} \overline{u}, &=& 0,
\end{eqnarray*}
which clearly has by solution  
\begin{equation}\label{eq:u}
    \overline{u} = -\dfrac{B}{r} + A,
\end{equation}
with constants $A$ and $B$. Since we want that $u \to 1$ as $ r\to \infty$, we demand that $A=0$.

Next, we move to the equations for the metric components. After a straightforward computation one finds after projecting the linearized static equations to the non-coordinate frame that $\overline{\delta(f R_{01})} = \overline{\delta(f R_{11})} = \overline{\delta(\nabla_{0} \nabla_{1} f)} =  \overline{\delta(\nabla_{1} \nabla_{1} f )} = 0$, where the covariant derivatives were computed by using the analog formula \Eqref{cov.derive}. This results is due to the fact that their spin-weight is nonzero and hence, they have a zero fundamental mode. As a result, we only have to solve the linearized static equations for the components $\mu,\nu=0,0$ and $\mu,\nu=1,2$. However, instead of solving the system  \Eqref{LEE} directly, we will solve the equations for the components $\mu,\nu=0,0$ and $\mu,\nu=1,2$ of
\begin{eqnarray*}
 \overline{ \delta f G_{\mu\nu} }   &=& \overline{ \delta ( \nabla_{\mu} \nabla_{\nu} f ) } \ ,   
\end{eqnarray*}
where 
\begin{eqnarray*}
    G_{\mu\nu}:= R_{\mu\nu} - \dfrac{ R}{2} g_{\mu\nu}
\end{eqnarray*}
is the Einstein tensor of $g$. Note that if $\overline{u}$ is solution of \Eqref{eq:u}, then the solutions of this system also must satisfy the later equation. 
Therefore, after a straight forward computation we obtain for the indices $\mu,\nu=1,2$:
\begin{eqnarray*}
 \overline{ \delta (f G_{12} ) }  &=& \overline{ \delta ( \nabla_{1} \nabla_{2} f ) } \ ,   \\
  r^2  \   \partial_{rr} \overline{h}  +  r    \partial_r \overline{h} &=& 2 r \ \partial_r \overline{u} \ .
\end{eqnarray*}
Using \Eqref{eq:u} we obtain
\begin{equation}\label{eq:h_ecuaction}
    \overline{h} =  \dfrac{2 B}{r} + C + D \ln{r} \ ,
\end{equation}
where $C$ and $D$ are integration constants. On the other hand, the equation for the indices $\mu,\nu=0,0$ gives
\begin{eqnarray*}\label{eq:constraint1}
 \overline{ \delta (f G_{00} ) }  &=& \overline{ \delta ( \nabla_{0} \nabla_{0} f ) } \ ,  \\
\frac{  \partial_r \overline{h} } {r} &=&   \partial_{rr}  \overline{u} \ .
\end{eqnarray*}
Therefore, if we substitute \Eqref{eq:h_ecuaction} and \Eqref{eq:u} we can easily see that it holds by choosing $D=0$. On the other hand, in order to find the constants $B$ and $C$,  we will use the boundary conditions as follows.

First, note that if we take the trace with respect to the metric of the two sphere $\mathring\gamma_{ }$ in the first boundary condition $ \overline{g}^{(\epsilon)}_{ }   =  \overline{\gamma}^{(\epsilon)}_{ }$ at $r=r_0$, we obtain
\begin{eqnarray*}
   2 \overline{h} \big|_{r=r_0} &=& \overline{ \text{tr}_{\mathring{\gamma}} \gamma}^{(\epsilon)},
\end{eqnarray*}
where $\overline{ \text{tr}_{\mathring{\gamma}} \gamma}^{(\epsilon)}$ is the fundamental mode of the trace of $\gamma^{(\epsilon)}_{ }$ with respect to the $\mathring{\gamma}$. 
Second, by using \Eqref{mean_curvature_1}, the second boundary condition can be written as 
\begin{equation*}
 \overline{\mathcal{H}}^{(\epsilon)}  = \left( -  \frac{ \overline{h}}{r}+  \partial_r  \overline{h} \ \right) \Big|_{r=r_0}= \overline{H}^{(\epsilon)}. 
\end{equation*}
Third, by substituting the solutions \Eqref{eq:h_ecuaction} and \Eqref{eq:u} in the two boundary conditions  leads to   the $2\times2$ algebraic system

\begin{eqnarray*}
     \dfrac{  \overline{ \text{tr}_{\mathring{\gamma}} \gamma} }{ 2 }  =  \dfrac{2 B}{r_0} + C, \quad 
      \overline H^{(\epsilon)}   =  \frac{ 4 B  }{r_0^2} + \dfrac{C}{r_0} ,
\end{eqnarray*}
which easily gives
\begin{equation*}
    B =  \frac{r_0}{2}  ( \overline{ H}^{(\epsilon)}   r_0 - \dfrac{ \overline{ \text{tr}_{\mathring{\gamma}} \gamma}^{(\epsilon)} }{2} ), \quad C =  \overline{ \text{tr}_{\mathring{\gamma}} \gamma}^{(\epsilon)}  - r_0 \overline{ H}^{(\epsilon)}.
\end{equation*}

After determining the constants, we can substitute them into \Eqref{eq:Mass_equation_final}. Then, utilizing \Eqref{minimizer}, we     obtain the Bartnik mass of $\Sigma$ in terms of the mean values of  $\overline{ \text{tr}_{\mathring{\gamma}} \gamma}^{(\epsilon)}$ and $\overline{ H}^{(\epsilon)}$ as

\begin{equation}\label{finalformulaADMmass}
  M_{B}(\Sigma) = \lim_{r \to \infty} \dfrac{r^2}{2} \left(  - \ \partial_r \ \epsilon \overline h \ \right) \approx B =  \frac{r_0}{2}  \epsilon \left( \overline{ H}^{(\epsilon)}   r_0 - \dfrac{\overline{ \text{tr}_{\mathring{\gamma}} \gamma}^{(\epsilon)}}{2} \right).
\end{equation}

\subsection{Consistency with Weygul's estimate}
 
We finalize this section by showing that the expression \Eqref{finalformulaADMmass} is consistent with Weygul's result \cite{wiygul2018bartnik}, which establishes that for a hypersurface $\Sigma$ that is close to $\mathbb{S}^{2}(1)$ with metric $\gamma_{ }$ and mean curvature $H$, the Bartnik mass is approximately equal to 
\begin{eqnarray*}
M_{B}(\Sigma)  \approx \dfrac{1}{16 \pi} \int_{\mathbb{S}^2(1)} (6 + 2  H  - \text{tr}_{\mathring{\gamma}} \gamma ) d \Omega^2 \ ,
\end{eqnarray*}
Writing the metric $\gamma_{ }$ and the mean curvature $H$ of $\Sigma$ as in \Eqref{bartnik_data}, and noting that $ \text{tr}_{\mathring{\gamma}} \mathring{\gamma} = 2$ and $\mathring{H}=-2$ for $r=1$, we can integrate the above  expression to obtain (see \Eqref{mean_value_functions})
\begin{eqnarray*}
    M_{B}(\Sigma) &\approx& \dfrac{1}{4} ( 6 + 2  \overline{H}  - \overline{ \text{tr}_{\mathring{\gamma}} \gamma }  ),\\
    &=& \dfrac{1}{4} (6 + 2 ( -2 + \epsilon \overline{H}^{(\epsilon)})  - ( 2 + \epsilon \ \overline{ \text{tr}_{\mathring{\gamma}} \gamma}^{(\epsilon)} ) ),\\
    &=& \dfrac{1}{4} ( 2 \epsilon \overline{H}^{(\epsilon)} - \epsilon \  \overline{ \text{tr}_{\mathring{\gamma}} \gamma}^{(\epsilon)}  ),\\
     &=&  \dfrac{\epsilon}{2} \left(  \overline{H}^{(\epsilon)} -  \dfrac{  \overline{ \text{tr}_{\mathring{\gamma}} \gamma}^{(\epsilon)} }{2} \right).
\end{eqnarray*}
Therefore, it is clear that \Eqref{finalformulaADMmass} is consistent with Weygul's expression who takes $r_0 = 1$. We want to remark that we have obtained \Eqref{finalformulaADMmass} by implementing a completely different approach (based on the eth-formalism) than the one used by Weygul for obtain his estimate. Furthermore, our formula generalizes Weygul's as it applies for $\Sigma$ close to 2-sphere of an \emph{arbitrary} radius $r_0$, $\mathbb{S}^2(r_0)$.

\section{Numerical construction of static metrics with some given Bartnik Mass}\label{sec:4}

When explicitly expressed in terms of the metric components and the scalar field $f$, the static system (\ref{EE}-\ref{BC}) defines a boundary value problem. Therefore, one would expect that it can be formulated as a system of elliptic partial differential equations. However, this is not the case. This becomes evident when considering the expression for the Ricci tensor, as illustrated in \cite{ringstrom2009cauchy}:
\begin{equation*}
R_{\mu\nu} = -\frac{1}{2} \Delta_{g} g_{\mu\nu } + \nabla_{(\mu} \Gamma_{\nu)} + H_{\mu \nu }(g, \partial g).
\end{equation*}
Here, $\Gamma_{\nu} := g_{\nu \mu } g^{\sigma \gamma} \Gamma^{\mu}_{\sigma \gamma}$ and $H_{\mu \nu }(g, \partial g)$ is a tensor dependent on the metric and its first-order derivatives. Furthermore, $ \nabla_{(\mu} \Gamma_{\nu)}$ is a function of $g,\; \partial g$ and $\partial \partial g$ and contains, in particular, second-order derivatives of the metric as well as the Laplacian $\Delta_{g} g_{ }$. Traditionally, if $\nabla_{(\mu} \Gamma_{\nu)}$ were zero due to a certain choice of coordinates (or replaced by another tensor without second-order derivatives of the metric), it can be shown that $R_{ }$ leads to an elliptic system of partial differential equations for the metric. This concept forms the basis of the 'harmonic gauge' used to establish the local existence and uniqueness of solutions to the Ricci-flow equation, as discussed in \cite{chow2004ricci}.

However, the ellipticity of the entire system is disrupted by the Hessian $\nabla_{\mu}\nabla_{\nu} f$. Consequently, in principle, standard numerical methods suitable for solving elliptic systems, such as a parabolic approach, cannot be employed. Hence, in this work, we propose an approach involving neural networks that, unlike standard numerical methods for boundary value problems, does not necessitate the system to be elliptic. For further insights into this subject, refer to the detailed review in \cite{yadav2015introduction} or the more recent and concise work presented in \cite{blechschmidt2021three}.

\subsection{The neural network approach}\label{sec:deep_learning}

Neural networks have become increasingly prevalent in constructing numerical solution methods for partial differential equations across various research domains. The primary reason for their popularity is the multitude of advantages they offer over conventional methods. For instance, neural networks excel at approximating highly non-linear and complex functions, enabling them to capture intricate relationships between variables. This capability is particularly advantageous for solving non-linear systems of partial differential equations. Therefore, in this work, we adopt this approach to solve the static system (\ref{EE}-\ref{BC}).

Following \cite{calin2020deep}, we can define a neural network as a computational model composed of interconnected nodes, also known as neurons or artificial neurons, organized into layers. Each neuron in a neural network is associated with an \textit{activation function} $\varphi$. The activation function introduces non-linearity to the model, allowing it to learn complex patterns.  The output of a neuron $y$, after applying the activation function, can be expressed as a weighted sum of inputs plus a bias term, followed by the activation function as
\begin{equation*}
    y=  \varphi \left( \sum_{i} \omega_{i} x_{i} + b \right),
\end{equation*}
where the set of $\omega_i$ represents the weights of the neuron, the $x_{i}$ the inputs and $b$ the bias term.  The layers, on the other hand, 
 are the structural components that organize all the the neurons as groups. The are three main types of layers, namely: the input Layers, hidden Layers and Output layers. The first type correspond to the input data of the Neural network. Each node in this layer represents a feature or attribute of the input. In our case this will be the independent coordinates of our manifold. The second type, the hidden layers, are the ones that contain  the different ensembles of neurons, the activation functions and the weights. Deep neural networks have multiple hidden layers, enabling them to learn complex representations of data. Finally, the output layers  produces the network's output or prediction.  In our case each neuron will correspond to a metric component, and thus, our neural network will lead to the function
  \begin{eqnarray*}
\mathcal{F} &:& \mathbb{R}^2 \rightarrow \mathbb{R}^5, \\
  & & (r, \theta,\phi) \mapsto (f,g_{11},g_{12},g_{22},g_{23}),
\end{eqnarray*}
where $\mathcal{F}$, usually called the \textit{learning function}, is a function composed of all the neurons of our neural network that depends of the complete set of parameters ${ \omega }$, ${ b }$  and activation functions. 
         
Training the neural network involves adjusting its parameters  to minimize a certain function usually called \textit{cost function}. This function is chosen based on the nature of the task that needs to be addressed (e.g., binary classification, multi-class classification, regression, etc). For our specific case, inspired by the work of \cite{lagaris1998artificial}, we propose the following cost function:
\begin{equation}\label{the_cost_function}
C(\mathcal{F}) : =   \lVert \Delta_{g} f \rVert^2 + \sum^{3}_{\mu=1} \sum^{3}_{\nu=\mu} \ \lVert f R_{\mu\nu } -  \nabla_{\mu} \nabla_\nu f \rVert^2,
\end{equation} 
where  $\mathcal{F}$ is the learning function. Hence, in principle, we aim to approximate solutions of the static system (\ref{EE}-\ref{BC}) by identifying the optimal combination of parameters and activation functions that minimize $C(\mathcal{F}) \approx 0$. To achieve this, we treat the cost function $C$ as a function of the parameters within the learning functions $\mathcal{F}$. We then employ a standard algorithm, such as the gradient descent method \cite{yadav2015introduction}, to search for the minimum of this function.

In what follows we will discuss some of the technical details of the  numerical implementation of this neural network  to the problem at hand, and we will present some numerical results.

\subsection{Choosing the form of the learning functions}\label{sec:learning_functions}


As mentioned earlier, our objective is to determine the Bartnik mass of a closed 2D hypersurface $\Sigma$ with a specified metric and mean curvature, as outlined in Equation \eqref{bartnik_data}. However, unlike the preceding section, we now assume that $\Sigma$ is not an almost round sphere, meaning it cannot be regarded as a linear perturbation of the metric and mean curvature of $\mathbb{S}^{2}(r_0)$. Mathematically, we express this by imposing the condition $\epsilon^2 \neq 0$ on the perturbation parameter.

From now on, we will  assume that the form of the perturbations will have the form 
 \begin{equation}\label{perturbation_final}
      g^{(\epsilon)}_{ } :=   \alpha \ \mathbf{w}^0 \otimes  \mathbf{w}^0  +  2 h r^2    \mathbf{w}^1 \otimes  \mathbf{w}^2.
\end{equation}
Note that this form of the perturbation differs from the one used in the previous section, see \Eqref{perturbationlinearregime}, on the assumption that the metric is not necessarily conformally flat.


In order to effectively solve the static system (\ref{EE}-\ref{BC}) of equations along with their corresponding boundary conditions, following \cite{lagaris1998artificial,cuomo2022scientific}, we must define a specific form for the learning functions. For simplicity we will assume that all metric components depend solely on the coordinates $r$ and $\theta$, implying axial symmetry. Furthermore, as our interest lies in evaluating the cost function $C$ across the entire domain, we will introduce the following coordinate transformation. This transformation is commonly utilized in finite element methods (refer, for example, to \cite{zienkiewicz1983novel}).
\begin{equation*}\label{ec:coordinate_transform_fem}
\xi := 1 - \dfrac{2 (r_0 - r_p)}{r - r_p},
\end{equation*}
where $r_p$, the pole, is some real value $r_p \in (-\infty,r_0)$. 
Then the unbounded interval $[r_0,\infty)$ is  mapped to the bounded interval $[-1,1]$. Furthermore, if we discretize $[r_0,\infty)$ in $N$ points $r_i$, for $i=0,...,N-1$, it induces a discretization on $[-1,1]$ of $N+1$ points $\xi_i = \xi( r_i )$ where 
\begin{equation*}
\xi_0 = \xi( r_0 ) = -1 , \quad \xi_{N-1} = \xi(r_{N-1}) , \quad \xi_{N} = \xi( \infty) =1.
\end{equation*} 
Note that the interval $[r_{N-1},\infty)$ is mapped to the bounded interval  $[\xi_{N-1},\xi_{N}]$. Since we are interested in solving the static system  with boundary conditions on the unit sphere,  we will choose $r_0=1$ and $r_p=0$. Thus, using the coordinate transformations
\begin{equation}\label{coordinate_transformation}
r \to 2/(1 - \xi),\qquad \theta \to \theta,\qquad \phi \to \phi,
\end{equation}
we can write the metric \Eqref{form_of_perturbations} as (see also \Eqref{perturbation_final})
\begin{equation}\label{metric:ecuacion_final}
g_{ } = \left(
\begin{array}{ccc}
 \frac{4 (1+\alpha )}{(\xi -1)^4}  & 0 & 0 \\
 0 & 0 & \frac{4 (1+h) }{(\xi -1)^2}  \\
 0 & \frac{4 (1+h)}{(\xi -1)^2}   & 0 \\
\end{array}
\right),
\end{equation}
where $\alpha$ and $ h$ are some functions of the coordinates  $\xi,\theta$. Note that when $  \alpha = h = 0$ we obtain the flat metric $\mathring g$ in the new coordinates. Therefore, we will search for solutions of the form
\begin{eqnarray}
           f   &=&  \epsilon \ \mathcal{F}_{0}, \label{learning_functions}  \\
           h   &=&  \epsilon \ \dfrac{(1-\xi)^3}{8}    \left[ \dfrac{(1+\xi)(1-\xi)}{2} \mathcal{F}_1 + B_1 \right] ,\\
           \alpha &=&  \epsilon \ \dfrac{(1-\xi)^3}{8} \left[ \dfrac{(1+\xi)(1-\xi)}{2} \mathcal{F}_2 + B_2 \right] .\label{learning_functions2}  
\end{eqnarray}
Here $\epsilon$ is some constant, the $\mathcal{F}_{i}$ with $i=1,2,3$, representing the components of the learning function
\begin{eqnarray*}
\mathcal{F} : \quad \mathbb{R}^2 \quad \rightarrow \quad \mathbb{R}^3, \quad \hspace{0.6cm} \\
  (\xi, \theta ) \mapsto (\mathcal{F}_1, \mathcal{F}_2, \mathcal{F}_3),
\end{eqnarray*}
where $B_1$ and $B_2$ denote functions that set the boundary values of the metric components based on the Bartnik data (\ref{BC}). When $\xi=1$ we have $\alpha = h = 0$, indicating that the metric $g_{ }$ becomes flat at infinity. Conversely, for $\xi=-1$, we have $h = B_1$ and $\alpha = B_2$. 

The specific form of the proposed learning functions will be justified later in this subsection, and the choices for $B_1$ and $B_2$ will align with the Bartnik data. Due to \Eqref{form_of_perturbations}, the metric on $\Sigma$ takes the form 
\begin{equation*}
 \gamma_{ } =\frac{4}{(\xi -1)^2}  \left(
\begin{array}{cc}
0       & (1 + h)  \\
(1 + h)  & 0       \\
\end{array}
\right).
\end{equation*}
Hence, for $B_1$, which determines the value of the metric function $h$ at the unit sphere, we have the freedom to choose it as needed. Moreover, analogous to the principles of the $3+1$ decomposition of spacetime metric (refer to \cite{alcubierre2008introduction}), we define the unit normal vector to this hypersurface as
\begin{equation*}
   n := ( 1/ \sqrt{ (1 + \alpha) / (\xi-1)^4 } , 0, 0 ).
\end{equation*}
Upon evaluating the mean curvature $\mathcal{H} = - \nabla_{\mu} n^\mu$ of the surface $\Sigma$ using the covariant derivative associated with the metric in  \Eqref{metric:ecuacion_final}, we obtain
\begin{equation*}\label{mean_curvature_formula}
  \mathcal{H} =   \frac{(\xi -1) \ \partial_\xi h - 2 (h +1)}{2 (\xi -1) (h +1) \sqrt{(\alpha +1)/(\xi -1)^4}} .
\end{equation*}
Therefore, by evaluating this expression at the inner boundary $\xi=-1$, we can express $\alpha$ in terms of the metric components $\gamma_{ }$ and the mean curvature $H$ as:
\begin{equation}\label{B2:form}
 \alpha \big|_{  \xi=-1}   = \left( \frac{2 ( \partial_\xi h + h +1)}{ H (h +1)} \right)^{2}_{\xi=-1} - 1 \ .
\end{equation}

Hence, we can utilize this expression to express the boundary function $B_2 := \alpha \big|_{ \xi=-1}$. It is important to note that this formula requires knowledge of $\partial_\xi \hat{h}$, which is not provided by the Bartnik data. However, during the implementation of the deep learning algorithm, obtaining $\partial_\xi \hat{h}$ becomes feasible since the learning functions $\mathcal{F}_1, \mathcal{F}_2$, and $\mathcal{F}_3$ are known from the beginning of the implementation with some random parameters. Therefore, by estimating a learning function for $h$ using given parameters, we can also compute its derivative $\partial_\xi h$ using automatic differentiation algorithms (refer, for instance, to \cite{baydin2018automatic}).

We finalize this subsection by pointing out that in the new coordinates $\xi$, the formula for the ADM mass   becomes
 \begin{equation*}
    M_{ADM}(g) = -2 \left(  \dfrac{\overline \alpha - \overline h}{ (-1+\xi)^3}  -  \dfrac{\partial_\xi \overline h}{ (-1+\xi)^2}   \right) . 
\end{equation*}
However, as our neural network-based algorithm aims to determine functions $\alpha$ and $h$ following the specific forms in \Eqref{learning_functions}, after substitution, we rewrite $M_{ADM} (g)$,  as
\begin{eqnarray}\label{eq:Mass_equation_final2}
  M_{ADM}(g) = \dfrac{\overline{B}_1 }{2}+\dfrac{\overline{B}_2 }{4},  
\end{eqnarray}
which by means of \Eqref{minimizer}, implies that $M_{B}(\Sigma)$  will solely depend on the boundary conditions of the metric and mean curvature. In our numerical implementation, we will utilize this simplified formula for determining $M_{B}(\Sigma)$. 

\subsection{Numerical implementation}

The numerical result that we will present in what follows were conducted by using the library \textit{Tensorflow}, which is an open-source framework   for building and deploying machine learning models using different programming languages. Visit \cite{TensorFlow} for a full review and tutorial on this software. 

\subsubsection{Computation of the cost function}
As mentioned in \Sectionref{sec:deep_learning}, training the neural network requires evaluating the cost function \Eqref{the_cost_function} across the entire numerical grid domain. To achieve this, we substituted the metric \Eqref{metric:ecuacion_final} into the static field equations \Eqref{SEE}. After a rigorous yet straightforward computation, we derived partial differential equations for the variables $\alpha$, $h$, and $u$. Specifically, we obtained seven equations: six from $f R_{\mu \nu } - \nabla_{\mu} \nabla_\nu f$ and one from $\Delta_{g} f$. These equations will be denoted as $E_0$ through $E_5$. 

Subsequently, we define a residual $\hat C_i := E_i$ for each of these equations, which measures the error in which $\alpha$, $h$, and $u$ do not satisfy the static system (\ref{EE}-\ref{BC}). Unfortunately, upon simple inspection of the expressions, it becomes evident that some of these equations contain factors of the form $(1-\xi)^{-2}$ (or $(1-\xi)^{-1}$), which poses a problem when evaluating the cost functions at $\xi=1$. To address this issue, we redefine the residual as
\begin{equation*}
C_i := E_i (1-\xi)^{2},
\end{equation*}
eliminating any problems at $\xi=1$. Consequently, we compute the cost function as
\begin{equation*}\label{the_cost_function2}
C(\mathcal{F}) : = \sum^{5}_{i=1} \lVert C_i \rVert^2,
\end{equation*}
where $\lVert \cdot \rVert$ represents the discrete maximum norm. For our spatial domain $[-1,1]\times[0,2\pi]$, we chose a discretization of $400$ points along the $\xi$ coordinate, while for the angular coordinate $\theta$, only $24$ points were necessary. 

As mentioned in \Sectionref{sec:learning_functions}, the spatial derivatives of the learning functions in the equation $E_i$ can be computed using the automatic differentiation algorithm. This algorithm essentially performs analytical differentiation of the learning function based on the known neural network architecture and then evaluates it at the grid points. For a comprehensive review of this algorithm, refer to \cite{margossian2019review}.


\subsubsection{Numerical results}

\begin{figure}[h]  
\centering
\includegraphics[scale=0.35]{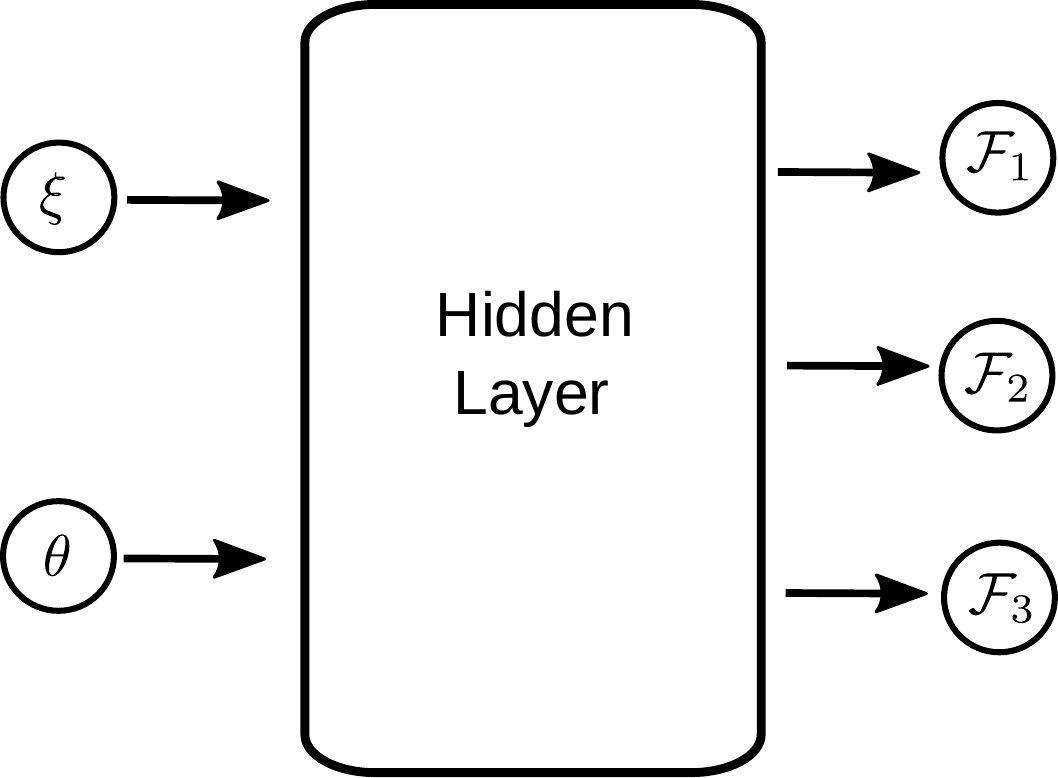}  
\caption{Scheme of the neural network}
\label{schemeNN}
\end{figure}

As mentioned in \Sectionref{sec:deep_learning}, the input layer consists of the coordinates, while the output layer corresponds to the metric components we aim to determine. In accordance with the learning function defined in \Eqref{learning_functions}, we selected two neurons for the input layer and three neurons for the output layer, without any activation function. Figure \ref{schemeNN} provides a simple illustration of the neural network utilized in our numerical implementation.

Subsequently, the determination of the number of layers, neurons per layer, and activation functions for the hidden layers of the neural network becomes crucial. Cybenko's seminal work in 1989 \cite{cybenko1989approximation}, along with subsequent studies (for a concise overview, refer to \cite{goodfellow2016deep}), established that even a single hidden layer network could serve as a universal approximator given a sufficiently large number of neurons. However, while the universal approximation theorem underscores this capacity, it doesn't offer precise guidance on choosing the network architecture or parameters like the number of layers, activation functions, or learning rate. Consequently, selecting these elements is pivotal as they profoundly influence the accuracy of the approximated functions.

Considering this, we experimented with various configurations until achieving satisfactory results based on the behavior of the cost function. Essentially, the objective was to explore different setups wherein achieving a lower value of the cost function post-training indicates a better configuration.

\begin{figure}[t]
    \begin{minipage}{0.47\linewidth}
        \centering
        \includegraphics[scale=0.5]{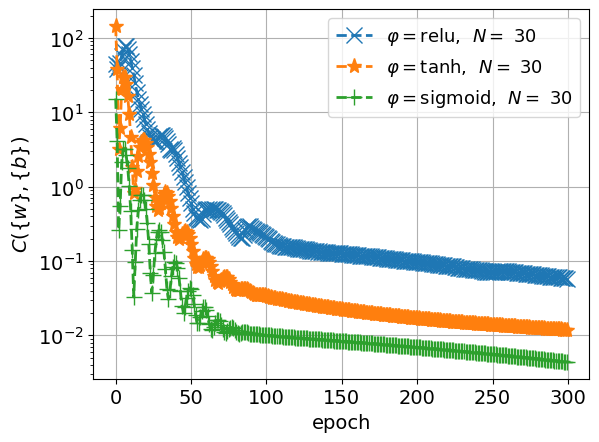}
        \caption{Behaviour of the cost function during training for different activation functions and a fixed number of neurons.} \label{fig:r11}
        \end{minipage}  
    \hfill
        \begin{minipage}{0.47\linewidth}
        \centering
        \includegraphics[scale=0.5]{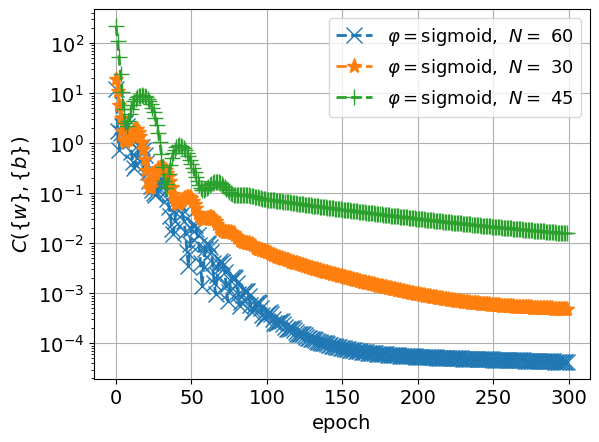}
        \caption{Behaviour of the cost function during training using the sigmoid activation function and a differing number of neurons.}\label{fig:r2}
        \end{minipage}  
\end{figure}

To gain insight into these questions pertinent to our specific problem, we aimed to determine the most suitable activation function for our neural network. For this purpose, we trained several neural networks with a single hidden layer comprising thirty neurons, each employing a different activation function. In Figure \ref{fig:r11},  we display the behavior of the cost function for three distinct activation functions: \textit{relu}, \textit{tanh}, and \textit{sigmoid}. We conducted training for our neural network over 300 epochs using the ADAM method, an optimization technique derived from the gradient method and readily available in software libraries (see, for instance, \cite{calin2020deep} for a comprehensive review of various activation functions and the ADAM method). Additionally, for this experiment, we set the following boundary conditions
\begin{eqnarray*}
    B_1 = \epsilon \sin^2 2 \theta, \quad B_2 = 0 .
\end{eqnarray*}


It is important to note that the second boundary condition should ideally be derived by first choosing $\mathcal{H}$ and then utilizing \Eqref{B2:form} to obtain $B_2$. However, due to practical reasons, in our implementation, we directly chose $B_2$ assuming that we had previously selected an appropriate $\mathcal{H}$. Additionally, for all subsequent experiments, we fixed $\epsilon = 0.5$.

In the figure, it is evident that the \textit{sigmoid} activation function (also known as the logistic activation function in the literature) exhibits the best behavior for the cost function. This could potentially be attributed to the fact that this type of function somehow mirrors the solutions we seek, displaying an asymptotic behavior as it approaches certain values within the domain.

Furthermore, in an attempt to determine an optimal number of neurons for our hidden layer, we trained various neural networks, each comprising a single layer with a different number of neurons, all employing the \textit{sigmoid} activation function. Figure \ref{fig:r2} illustrates three experiments conducted with $30$, $45$, and $60$ neurons. Surprisingly, the number of neurons in the layer does not necessarily ensure a better behavior for the cost function. A notable observation can be made by comparing the cases with $30$ and $45$ neurons, where the cost function displayed more favorable behavior in the former case than in the latter.
\begin{table}
    \centering  
    \begin{tabular}{cc}      
    \includegraphics[scale=0.60]{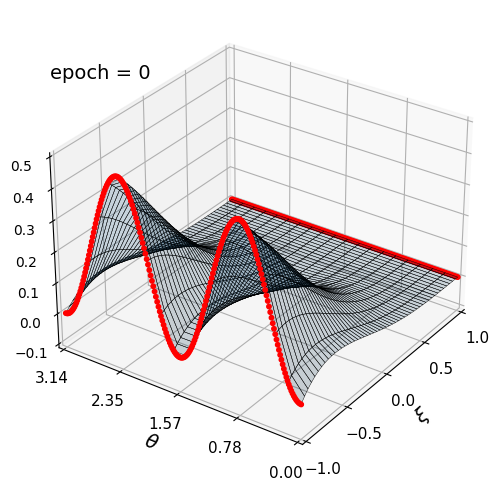}  &  \includegraphics[scale=0.60]{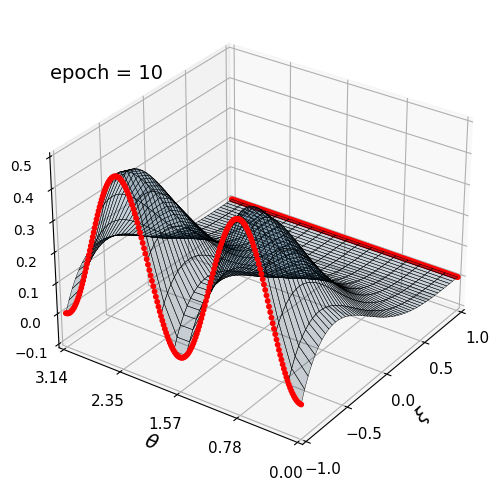}\\
    \includegraphics[scale=0.60]{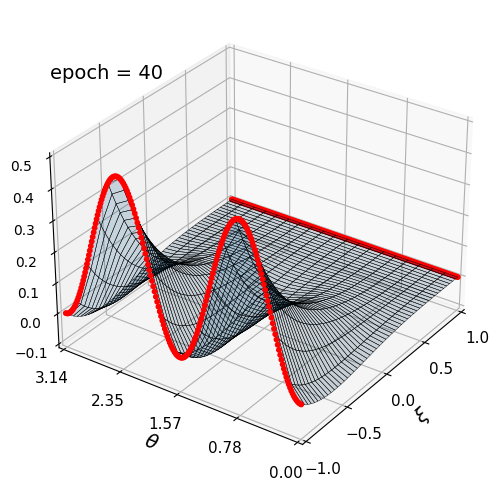} & \includegraphics[scale=0.60]{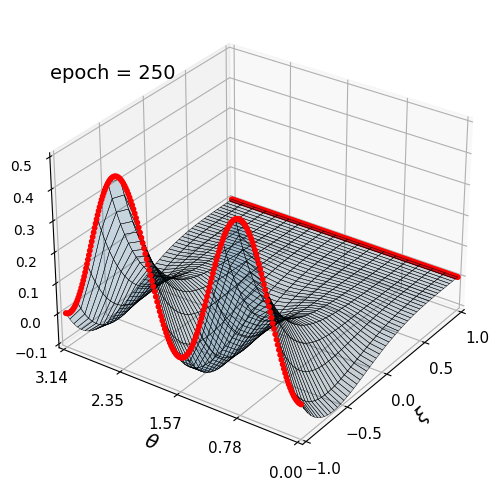}\\         
    \end{tabular}
    \caption{Sequence of for the behaviour of the function $h$ along the training of the neural network. The red color indicates the form of $h$ at the boundaries $\xi=-1$ and $\xi=1$ respectively.}
    \label{tbl:table_of_figures}
\end{table}

In the figures presented in Table \ref{tbl:table_of_figures}, we depict the evolution of the function $h$ throughout the neural network's training process. Specifically, these figures showcase the boundary conditions at $\xi=-1$ and $\xi=1$ for this function, that is: 
\begin{equation}
    h|_{\xi=-1}= B_1, \quad h|_{\xi=1}= 0.
\end{equation}
Qualitatively, we observe that the most significant changes manifest along the $\xi$-direction, whereas changes along the angular direction $\theta$ are imperceptible due to the scale. Notably, this behavior has been consistently observed across all experiments conducted with various neuron counts and activation functions.

\begin{table}
    \centering
    \begin{tabular}{cc}      
    \includegraphics[scale=0.60]{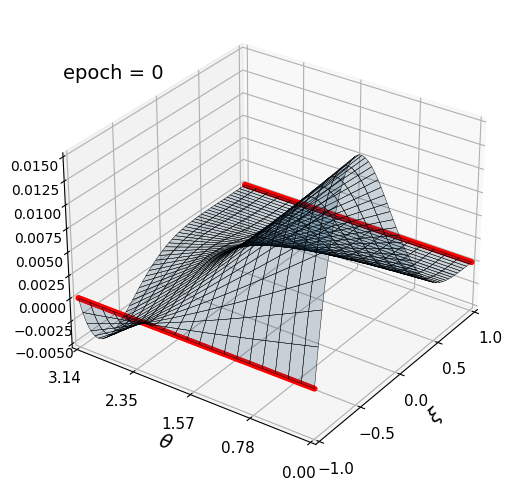}  &  \includegraphics[scale=0.60]{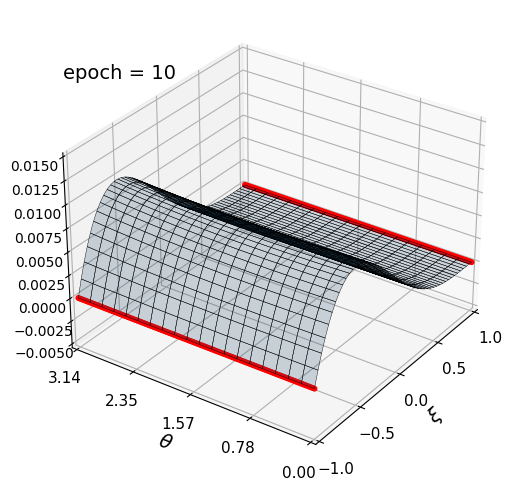}\\
    \includegraphics[scale=0.60]{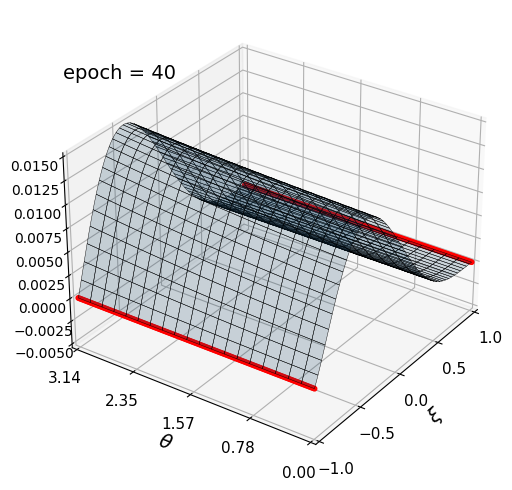} & \includegraphics[scale=0.60]{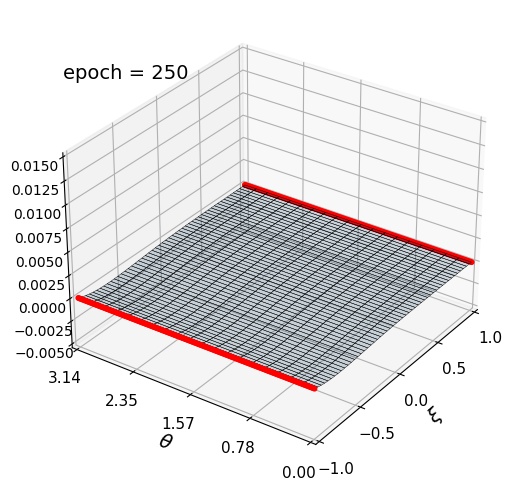}\\         
    \end{tabular}
        \caption{Sequence of for the behaviour of the function $\alpha$ along the training of the neural network.  The red color indicates the form of $\alpha$ at the boundaries $\xi=-1$ and $\xi=1$, that is $\alpha|_{\xi=-1}=B_{2}= \alpha|_{\xi=1}=0$.}
    \label{tbl:table_of_figures2}
\end{table}


For completeness, we also depicted the behavior of the variables $\alpha$ and $u$ in the Tables of Figures \ref{tbl:table_of_figures2} and \ref{tbl:table_of_figures3}, respectively. Note, however, that unlike the quantities $h$ and $\alpha$, $u$ does not have any boundary condition, and thus, its behavior diverges from the other two in that it resembles a plane being progressively elevated throughout the training phase of the neural network.


To conclude this section, we provide some remarks on the numerical solution we have just obtained. Firstly, owing to the form of the learning functions (see Equations \eqref{learning_functions}–\eqref{learning_functions2}), the numerical solution exhibits asymptotic flatness. Specifically, as $\xi \to 1$ (corresponding to $r \to \infty$ in the standard radial coordinate, as indicated by the coordinate transformation in Equation \eqref{coordinate_transformation}), the metric $g$ tends towards the flat metric. This is evident because the fields $h$ and $\alpha$ vanish. The asymptotic flatness is crucial for facilitating the application of the formula \eqref{eq:Mass_equation_final} in approximating the Bartnik mass.

Secondly, owing to the form of the numerical solution, the Bartnik mass $M_{B}$ for these solutions is entirely determined by the mean values of the boundary conditions $B_1$ and $B_2$ (refer to Equation \eqref{eq:Mass_equation_final2}). However, in our specific choice of setting $B_2$ to zero, we have obtained an approximate numerical solution where $M_{B}$ depends solely on the mean value of the perturbation of the metric on $\mathbb{S}^2(1)$. In simpler terms, in this scenario, the Bartnik mass is dictated by half of the mean value of the perturbation in the radius of a unit two-sphere. It is crucial to note that in order to maintain $M_{B}$ as positive, it is required that this mean value be positive. Moreover, in the general case, the choices of $B_1$ and $B_2$ must be made to ensure the satisfaction of this condition.

\begin{table}
    \centering
    \begin{tabular}{cc}      
    \includegraphics[scale=0.60]{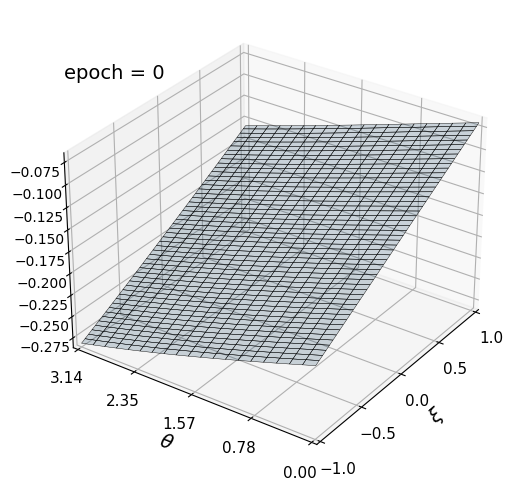}  &  \includegraphics[scale=0.60]{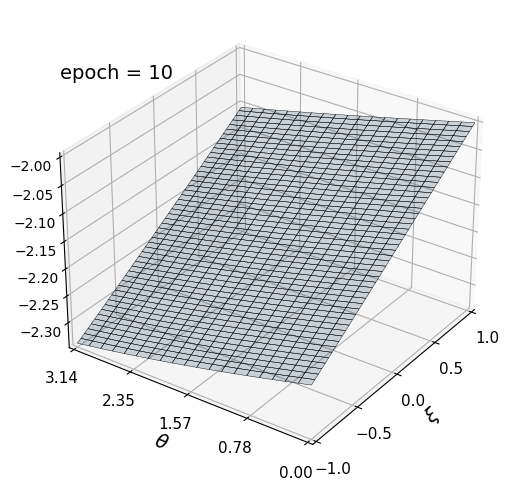}\\
    \includegraphics[scale=0.60]{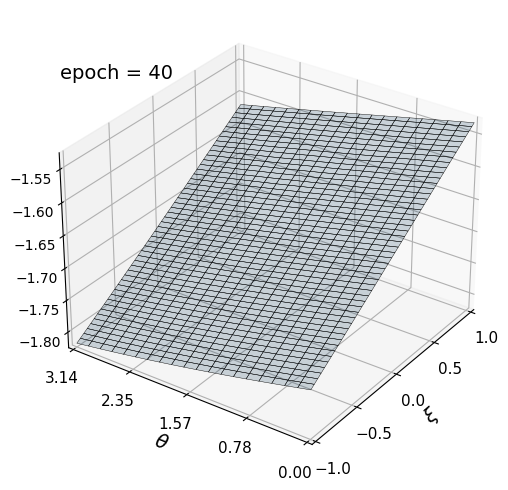} & \includegraphics[scale=0.60]{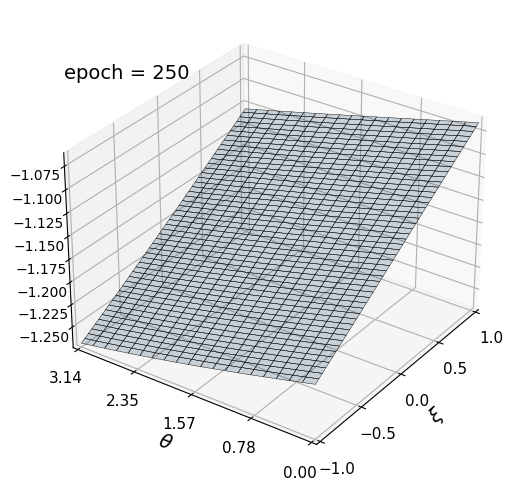}\\         
    \end{tabular}
        \caption{Sequence of for the behaviour of the function $u$ along the training of the neural network.}
    \label{tbl:table_of_figures3}
\end{table}


\section{Discussion}\label{sec:5}

In this study, the eth-formalism has been employed to represent $M_{ADM}(g)$ of a static space-time metric $g$ by averaging its components over a two-sphere, and we found that only metric components with zero spin-weight contribute to the ADM mass.  Although this expression is initially derived for standard spherical coordinates,  we highlight its broader applicability in expressing the Bartnik mass $M_{B}(\Sigma)$ associated with a surface $\Sigma$.

In the initial part of this study, we derived an expression for $M_{B}(\Sigma)$ in the scenario where $\Sigma$ is a linear perturbation of $\mathbb{S}^{2}(1)$, as indicated by \Eqref{finalformulaADMmass}. Furthermore, we validated its consistency with Wiygul's estimate for the unit two-sphere, as referenced in \cite{wiygul2018bartnik}, and extended it to a two-sphere of unspecified radius. 

We firmly believe that this expression holds significant potential for various applications. For instance, it can be utilized to calculate the Bartnik mass of a gravitational wave interacting with a black hole, enabling comparisons with the globally Bondi mass of the spacetime computed at future null infinity, as demonstrated in \cite{frauendiener2021non, frauendiener2023non, frauendiener2023non2}. 

On the other hand, Neural networks are emerging as potent tools for tackling diverse numerical challenges in general relativity, thanks to the advantages they offer over traditional methods (see, for instance, \cite{luna2023solving} and references therein). Thus, in addition to the above, in the latter part of this investigation we introduced a neural network methodology aimed at numerically constructing static metrics that incorporate these 2D-hypersurfaces while adhering to specified Bartnik masses. The application of this approach has enabled the successful approximation of numerical solutions for the static system (\ref{EE}-\ref{BC}). These outcomes underscore the promising potential of employing this methodology to provide solutions for various scenarios within gravitational physics or related fields. For instance, this approach could be readily applied to consider purely radiative spacetimes, facilitating investigations into the quasi-local energy contained in gravitational waves.

We conclude by noting that, to the best of our current knowledge, there exists another noteworthy numerical proposal designed to address this issue, centered around an inverse mean curvature geometric flow (refer to \cite{Cederbaum2019AFA}). However, while this method demonstrates remarkable accuracy in numerically solving the static system (\ref{EE}-\ref{BC}), it is confined to 2-dimensional axial symmetric hypersurfaces due to its reliance on the Weyl–Papapetrou formalism for axisymmetric static solutions of the Einstein vacuum equations. In contrast, although our approach may be less precise, the neural network methodology presented here for solving the boundary value problem offers the flexibility to approximate the Bartnik mass across a broader spectrum of 2D-hypersurfaces. Moreover, from a technical standpoint, computationally intensive boundary value problems solved using traditional CPU-based solvers may find advantages in such a scheme, naturally inheriting the computational benefits associated with GPU-based codes.



\section*{Acknowledgments}
This work was supported by Patrimonio Autónomo - Fondo Nacional de Financiamiento para la Ciencia, la Tecnología y la Innovación Francisco José de Caldas (MINCIENCIAS - COLOMBIA) Grant No. 110685269447 RC-80740-465-202, projects 69723 and 69553.

\addcontentsline{toc}{section}{References}
\bibliographystyle{abbrv}
\bibliography{references}

\end{document}